\documentstyle[12pt]{article}

\def\be{\begin{equation}}
\def\ee{\end{equation}}
\def\a{\alpha}
\def\d{\delta}
\def\D{\Delta}
\def\g{\gamma}
\def\o{\omega}
\def\O{\Omega}
\def\s{\sigma}

\def\s{\sigma}
\def\e{\epsilon}
\def\f{\varphi}
\def\z{\zeta}
\begin{document}
\vspace{1.5in}
\begin{center}

{\large \bf  Band splitting and relative spin alignment
in bilayer systems.} \\

\vspace{0.4in}
{ A.A.~Ovchinnikov$^{1)}$ and M.Ya.~Ovchinnikova$^{2)}$. }   \\
\vspace{0.2in}
{\it  $ ^{1,2)}$Joint Institute of Chemical Physics, RAS, Moscow.} \\
{\it  $ ^{1)}$Max Planck Institute for Physics of Complex Systems, Dresden.}
\\ \vspace{0.2in} \end{center}

\begin{abstract}

Influence of relative spin alignment on the band splitting and magnetic
excitations  in bilayer correlated systems is studied. Splitting occurs
to be large or small depending on relative orientation of staggered
spins of the layers.  Change of the ground state spin configuration with
doping is shown.  Behavior of bilayer splitting in $Bi_2Sr_2CaCu_2O_{8+\d}$
allows to suppose that superconducting transition is accompanied by
transformation of spin configuration of system.

\end{abstract}

\vspace{0.12in}

PACS: 71.10.Fd,  71.27.+a, 71.10.Hf

\vspace{0.12in}

One of properties  of high $T_c$ superconductors - a dependence of their
characteristics on number of the $CuO_2$ layers - shows importance of
interlayer coupling in them. In recent photoemission experiments \cite{1,2,3}
the bilayer splitting of the bands and the Fermi surfaces (FS) in
$Bi_2Sr_2CaCu_2O_{8+\d}$ (BSCCO) have been observed. Unusual observation was
a significant decrease in bilayer splitting at $k\sim (\pi,0)$ in course of
the superconducting (SC) transition \cite{2}. Another phenomena which is
observed in the bilayer (but not monolayer) cuprates is the appearance in
SC state of a spin resonance in inelastic neutron scattering
\cite{4,5,6}.
Explanation of these phenomena is important for understanding the electronic
structure and mechanisms of HTSC displaying  many properties of doped Mott
insulator. The main feature of latter is the Hubbard splitting of band
induced by antiferromagnet (AF) fluctuations. Though the long range AF order
disappear at small doping $\d_c=1-n_c\sim 0.05$, the local 2D spin order
probably retains in sufficiently large doping range including the range of
superconductivity. Among arguments in favour of such local AF order
there are  data on $\mu$SR, NMR, NQR \cite{7,8}, large length of spin
correlations with $Q=(\pi ,\pi )$ and smooth evolution of collective magnetic
excitations with doping (see review \cite{4}) and others. Finally, recently
a direct proof of the AF order even in SC state of
$YBa_2Cu_3 O_{6.5}$ have been obtained \cite{9,10} in nanosecond scale.

Detailed analysis of the AF bands and interactions of the Hubbard model have
been done in classic work \cite{11}. Simple mean field method gives
the overestimated values $d_0$ of staggered spin and of the doping boundary
$\d_c\sim 0.45$ of a range of the AF order. The band renormalization on base
of the zero AF approximation \cite{11} or the calculations of t-t'-U
Hubbard model using the slave-boson technique \cite{12} or the states with the
valence bond correlations \cite{13} - all these methods give the smaller
values of $d_0$ and of the boundary doping
$\d_c\sim0.3$ at which the local magnetization disappears. According
\cite {13} the range $\d<\d_c$ includes entirely the range of
superconductivity.  Retaining of the Hubbard splitting of band at $\d<\d_c$
in the t-t'-U \cite{13} and t-t'-J \cite{14,15,16} models implies the change
of the FS topology at the optimal doping with transition from the hole
pockets to large FS.

If such picture is true then interactions between the layers in bilayer
systems may significantly depend on relative orientation (alignment) of local
staggered spins of the layers. It must be so even if the spin orientation is
not stationary in strong sense, but is retained only on some time interval.
In case of small difference in the energies of various configuration of spin
alignment one can expect a strong influence of doping and temperature on
the properties of magnetic excitations in system. A search of such effects
may help to elucidate the origin of the magnetic resonance in neutron
scattering in YBCO, BSCCO.

Present work summarizes the results of variational calculations of bands,
full energies, phase curves $T_c(\d )$  and the band splittings in the
bilayer t-t'-U Hubbard model. Calculations are made on base of the correlated
states with the valence bond (VB) formation - the band analog of the RVB
state of Anderson.  Previously \cite{13} it was shown that the correlated
hopping interaction appearing in the effective Hamiltonian during the VB
formation does provide the hole attraction in d- channel and corresponding
d-superconductivity compatible with AF spin order. Here we use the same
method for study the effects of interactions of two layers. The questions of
interest are: Is the large doping range of local AF order retained? If so,
then what are the relative alignment of spin systems of both layers and the
bilayer splitting at various doping?

A difference between given approach and the study \cite{17} of bilayer
Hubbard model by the FLEX (fluctuation exchange) method must be stressed.
According FLEX the hole attraction is provided by long range spin
fluctuations. In our approach the short range AF correlations are decisive.
In simpliest variant they correspond to the VB formation, i.e. formation of
singlets on the bonds of the nearest centers. Unlike the averaged
consideration in FLEX we make the calculations for two specific spin
configurations of bilayer system - homogeneous states with the same or
opposite alignment of staggered spins of both layers. We denote them as the
$F_z$ or $AF_z$ (ferro- or antiferro-magnet over z direction) alignment
correspondingly, though both cases refer to local AF spin order in each
layer.  We consider only homogeneous states, though many neutron experiments
give indication on the stripe-type or spiral incommensurate structures in
cuprates. Nevertheless the effects discovered in our calculations for two
homogeneous spin structures must undoubtedly be taken into account in
discussing the observed bilayer splitting and magnetic properties of cuprates.

We start from the Hubbard Hamiltonian of bilaeyr system $H(U,t,t',t_z)$.
The main intralayer interactions are determined by
standard parameters $U,t$ of Hubbard model. Additional interactions include
the hoppings $t', ~t_z$ between the next nearest centers inside the layers
and between the cites of different layers. Positive signs of standard
parameters of strong coupling $t,~t',~t_z$ are defined in such way that the
zero bands (at $U=0$) have a form
\be
\e_{\pm}^{(0)}=-2t(cos{k_x}+cos{k_y}) +4t'cos{k_x}cos{k_y}
\mp {1\over 4}t_z(cos{k_x}-cos{k_y})^2
\label{1}
\ee
Expression for splitting $\D\e (k)=\e_+ -\e_- $  between the bonding and
antibonding zero bands is derived in \cite{17}.

A variational correlated state $\Psi=\hat{W}(\a)\Phi$ with correlations of
the VB type is constructed \cite{13} as unitary transformed uncorrelated
state $\Phi$
\be
\hat{W}(\a)=\exp\Bigl[\a Z\Bigr]; \quad Z= {{{1\over 2}}}
\sum_{\s,<nm>,\g}(c_{\g n\s}^\dagger c_{\g m\s}-h.c.) (n_{\g n,-\s}-n_{\g
m,-\s});
\label{2}
\ee
Here $\g=0,1$ refer to layers a,b.
A choice of unitary operator $W(\a)$ with variational
parameter $\a$ is explained in \cite{13}. Variational treatment of
problem with Hamiltonian $H$ in basis of correlated state $\Psi$ is
equivalent to treatment of the effective Hamiltonian
$H_{eff}(\a)=W^{\dagger}(\a)HW(\a)\approx  H +\a [H, Z]
+ {\a^2\over 2} [[H, Z]$
in mean field approximation. We use the most general uncorrelated states
$\{\Phi\}$ of the BCS type with anomalous averages of d-symmetry and the AF
spin order. The mean energy over such state
$<H>_{\Psi}=<H_{eff}>_{\Phi}=\bar{H}(y_{\g\nu},z)$
is obtained as an explicit function of a set of one-particle averages
$\{y_\nu , z\}$ over uncorrelated state $\Phi$. The set of these averages
contains the density components $r_{\g l}=<c_{\g n\s}^\dagger c_{\g,n+l,\s}>$,
analogous components of staggered spin
$d_{\g l}=(-1)^n {\s\over \mid\s\mid}<c_{\g n\s}^\dagger
c_{\g ,n+l,\s}>$, anamalous averages
$w_{\g l}=(l_x^2-l_y^2){<c_{\g n\s} c_{\g,n+l,\s}>}$ of d-symmetry for each
layer $\g =0,1$ and a value $z=t_z^{-1}<T_z>$ determining the average
interlayer hopping. For spin components we study two variants of relative
alignment $d_{\g =1,l}=\zeta_d d_{\g=0,l},~~\zeta_d=\pm 1$.
Selfconsistent procedure of the energy minimization over $\Phi$ and
of subsequent minimization over $\a$ is standard one \cite{13}. Details of
calculations will be presented elsewhere.

For main parameter we use a value $U/t=8$ following from
cluster derivation of single band model. Parameter $t'$ owing to its
contribution into the band dispertion  regulates the position
$E_{VHS}=E(\pi,0)$ of the van-Hove singularity (VHS) in the density of
state relative to the edge of the lower Hubbard band. In the monolayer
$t-t'-U$ or $t-t'-J$ models a value of $t'$ is directly connected with the
optimal doping $\d_{opt}$, since a maximum of $T_c$ corresponds to
coincidence of the chemical potential $\mu$ with $E_{VHS}$. According the
models \cite{13,14,15,16}  a resonable value of
$\d_{opt}\sim 0.2\div 0.25$ corresponds to values $t'/t\sim 0.05\div 0.1$.
These values are less than $t'/t\sim 0.2\div 0.4$ obtained from
fitting of photoemission data and the band calculation data on base of
the strong coupling scheme.  Here we vary parameter $t'$ in
limits $t'=0.05 \div 0.3$. Parameter of interlayer hopping is also varied
as $t_z/t=0.07\div 0.3$. Previous calculations without renormalization gave
$t_z/t\sim 1/3$ \cite{17}.  Recent measurements of bilayer splitting
$\D=2t_z=\d\e_k(\pi,0)$ \cite{1,2} give the value $t_z/t\leq 0.1$.

Fig.~1 presents typical doping dependencies of spin density $d_0$,
transition temperature $T_c(\d)$ and a difference
$\D H={\bar H}(AF_z)-{\bar H}(F_z)$ of average energies of bilayer system
with two types of relative alignment of the layer's spins. Staggered spin
density $d_0$ has nonzero value at wide range of doping including the region
of supercondutivity. Quantity $\D H(\d)$ changes its sign at some doping. In
the undoped system, $\d=0$, the
configuration $AF_z$ with opposite alignment of the
layer's spins occurs to be lower than the $F_z$ configuration. At large
doping $\d$ one have inverse situation  up to  boundary value $\d_c$, at
which the local magnitization disappears. Negative sign of $\D H$ at
$\d=0$ is explained by a positive exchange constant $J_{ab}\sim 2t_z^2/U>0$.
Change of sign of $\D H$ at some $\d$ is a result of the bilayer splitting of
the bonding and antibonding bands, in particular, the splitting of VHS, and
different occcupancy of them for two types of spin alignment.
Maximum of $\D H(\d )$ corresponds to optimal doping $\d_{opt}$
for the parent monolayer model. At this doping one has
$E_{VHS}^{1L}-\mu=0$. Increase of $t'$ shifts both the  $\d_{opt}$
and the position of maximum  of $\D H(\d )$. Such relation is not occasional.

According \cite{17} the bilayer splitting
$\d\e^{(0)}_z=2t_z(\cos{k_x}-\cos{k_y})^2/4$
of the zero bands has a maximum in region $k\sim (\pi ,0)$ forming the VHS
in density of state. However, the splitting contributes to the average
energy only if at this $k$ bonding and antibonding bands appear to be
occupied and unoccupied correspondingly. The latter takes place when
$E^{1L}(\pi,0)=E_{VHS}^{1L}=\mu$ for the unsplit bands of the monolayer model.
It remains to explain why a bilayer splitting manifests itself in the energy
difference $\D H={\bar H}(AF_z)-{\bar H}(F_z)$. The reason is in different
bilayer splitting for different spin configurations.

Fig.~2 presents the density of state of lower Hubbard band for two
configurations of spin alignment for system with small values $t',~t_z$.
Bilayer splitting is absent at $AF_z$ alignment, but sharply seen in $F_z$
configuration. This is due to different behaviour of matrix elements of
the interlayer hopping between the states of lower Hubbard band of each layer.
At $AF_z$ alignment these matrix elements disappear on the nesting lines.
Different behaviour of DOS reflects itself directly on the form of the phase
curves $T_c(\d)$ for SC transition (Fig.~1). In case of $AF_z$ or $F_z$
configuration the curve $T_c(\d)$ has  one or two maxima correspondingly.

Thus the models with small $t',~t_z\mathop{_\sim^<} 0.1$ are characterised
by 1) large $T_c^{max}\sim 0.02t\sim 116K$ at $AF_z$ configuration; 2) close
energies of both configurations, 3) small condensation energy; 4) lower $F_z$
configuration at $\d\sim \d_{opt}$; 5) large (or zero) bilayer splitting
in $F_z$ (or $AF_z$) configuration.

Unlike the normal state, a prediction of a lower spin configuration in
the SC state is hardly possible. An estimate of the condensation
energy and the step in a heat capacity occurs to be less than that observed
in YBCO approximately in 5 times. For this reason we dicuss the spin
configuration of SC state on base of the observed behaviour of bilayer
splitting in BSCCO \cite{1,2}. This splitting decreases from $\d\e\sim 80
meV$ in normal state down to $\d\e\sim 20 meV$ in SC state. Such behaviour
might be explained if we suppose that a transformation of configuration
$F_z\to AF_z$ proceeds simultaneously with SC transition.  One might suppose
also that the SC transition as such is induced by the change of
configuration, since such transition is accompanied by increase of the
density of states at the Fermi level.

For models with large values $t',t_z$ the types of dependencies $\D H(\d )$,
$d_0 (\d )$ are retained. Increase of $t_z$ leads to increase of $|\D H|$ and the doping
of the $F_z~-AF_z$ crossover. Calculations confirm the wide doping range of
AF local spin order for these model also. Value of $d_0$ is greater by an
order of magnitude than the value of AF spin density observed in SC state of
YBCO \cite{9,10} from elastic neutron scattering. The difference, possibly,
is connected with large distribution of the spin directions (or signs) of
different biplaines in crystal. Significant decrease of $T_c$  and a
deformation of the phase curves with increase of $t',~ t_z$ are connected
with decrease of the density of state on the FS for these models.

Influence of the spin alignment on the magnetic excitation spectrum
is of interest in connection with discovery of the spin resonance in neutron
scattering in SC state of cuprates and its unusual dispersion \cite{10}.
A widely discussed hypothesis \cite{18} connects its origin with the so
called $\pi$-resonance - an excitation of $e-e$ pair with $Q\sim (\pi,\pi )$
in triplet state. But the ratio of weights of  such resonance to the integral
(over $\o$) intensity of the non-resonance
magnetic excitations would be too small.
Expected order of magnitude of this ratio is $\sim (w/d_0)^2\sim 0.01$.
Interpretation of incommensurate patterns in the spin
susceptibility on base of the
nesting properties of FS also cannot describe a large intensity of spin
excitations and their similarity in both the undoped dielectric materials
and the doped metallic ones. The most probable hypothesis \cite{19} implies
the common magnetic origin of spin resonance  and of the incommensurate
anomalies of spin susceptibility $\chi ''(q,\o )$ in cuprates. Both features
are explained by existence of the AF domains or the incommensurate modulation
of the local staggered magnetization. The stripe-type structures have been
observed in $LaSrCuO$ and they may also present in the bilayer cuprates. But
monolayer models \cite{19} cannot explain the fact that the resonance
appears only in superconducting state and only in odd channel of bilayer
system. In this connection it is important to study the influence of spin
alignment on magnetic excitations in bilayer systems.

Here we study such effects using a simpliest example of homogeneous model
for "constructing" the spin resonace. Consider the fenomenological spin
Hamiltonian with anisotropic interlayer interaction
\be
H=\sum_{\g =a,b}J_0\sum_{<nm>}{\bf S}^{\g}_ n{\bf S}^{\g}_ m+
\sum_{n}\{ J_{z,ab}{S}_{zn}^a S_{zn}^b
+J_{\perp,ab}[S_{xn}^a S_{xn}^b+S_{yn}^a S_{yn}^b ]\}
\label{3}
\ee
The "$z$"-direction is selected by vector of local staggered spins $<{\bf
S}^\g_n>={\bf e}_z (\z_d)^\g d_0$  of each layer ($\g=0,~1$ for layers a,b)
with definite alignment $\z_d=1$ or $-1$, stabilized by the nonspin
interactions. Then the RPA consideration in the linear spin wave theory
\cite{20} gives the following frequencies of collective spin excitations
for both types of alignment
\be
\begin{array}{lll}
F_z:&~~~\o^{even}(q)\approx 2\O\sqrt{\f-g_1},&~~
\o^{odd}(q)\approx 2\O\sqrt{\f -g_2};~~~~\\
AF_z:&~~~\o^{even}(q)\approx 2\O\sqrt{\f+g_2},&~~
\o^{odd}(q)\approx 2\O\sqrt{\f +g_1};~~~       \\
\end{array}
\label{5}
\ee
Here $\f=2+\cos{q_x}+\cos{q_y}$, $\O=2J_0d_0$,
$g_{1(2)}=(J_{z,ab}\mp J_{\perp , ab})/2J_0$.

Fig.~3 presents a dispersion of the spin excitations in even and odd channels
for both configurations of a spin alignment for model with parameters
$g_1=0.05$, $g_2 =0.15$. Consider them in light of the above suggestion
about simultaneous transformation $F_z\to AF_z$ of the spin configuration
and superconducting transition in bilayer cuprates. If it is true for our
spin model (3), then at $T<T_c$ the intense peak appears in the odd channel
of $\chi^{odd}$ at $q\sim (\pi,\pi)$ with the frequency
$\o=\D^{odd}=2\O\sqrt{g_1}$, equal to a gap in the excitation spectrum in this
channel for $AF_z$ spin configuration. This peak is absent in case of $F_z$
configuration which is supposed to be ground state configuration at $T>T_c$.
Note that for $F_z$ configuration the low frequency excitations in
$\chi ''(q,\o\to 0)$ correspond to momenta $q$ distributed along the circle
$|{\bf q}-{\bf Q}|=\sqrt{2g_1}$ instead of  the discrete incommensurate
momenta $Q_{\eta}=(\pi (1+\eta ),\pi )$, observed for the low frequency
excitations in YBCO \cite{13}.They imply an existence of the incommensurate
spin structures. Fig.~3c schematically presents the branches of the spin
excitations for the spiral states or the states with modulation of a local
spin {$<S_{zn}>=d_0\cos{{\bf Q}_\eta {\bf n}}$}
of the monolayer (1L) model \cite{20}.
In \cite{20} the resonance frequency have been identified as a frequency at
the crossing point of branches at $q=(\pi ,\pi )$.

Thus, it is confirmed that local magnitization $d_0$ is retained in large
doping range for wide diapason of parameters $t',~t_z$. The crossover of two
spin configurations of the bilayer system proceeds at some doping. The lower
normal state is that with the $AF_z$ or $F_z$ spin alignment at small or large
doping correspondingly.  Maximum of the energy difference of these
configurations corresponds to the optimal doping of the parent monolayer
system and is connected with the maximum splitting of van-Hove singularities.
The bilayer splitting is zero or small for the $AF_z$ configuration, but
large for the $F_z$ spin alignment.  The observed in BSCCO large bilayer
splitting in normal state, but small splitting in SC state \cite{2} may
indicate according our models that the transformation $F_z\to AF_z$ of
configuration proceeds simutaneously with supercunducting transition. It is
shown that the magnetic excitation spectrum depends dramatically on type of
a spin configuration. The model spin system is presented for which a change
of configuration $F_z\to AF_z$  is accompanied by appearance of resonance (or
spin gap) in $\chi^{odd}$ at $q=(\pi,\pi )$.

A check of the above hypothesis requires further study, in particular
calculations of magnetic excitations in case of the inhomogeneous (of the
stripe type) spin structures of bilayer systems.

Work is supported by Russian Fund of Fundamental Reasearch (Projects
No. 00-03-32981 and No. 00-15-97334. Athours thanks V.Ya Krivnov for useful
discussions and P.Fulde for possibility to work in
Max Planck Institute for Physics of Complex Systems, Dresden.

\vspace{0.2in}


\vspace{0.2in}

{\bf Captions to Figures}
\vspace {0.15in}

Fig.~1.
Doping dependence of the local magnitization $d_0(\d )$ (top), of the
$T_c(\d )$ (middle) and value $w_1(\d)$ of the anamalous order parameter
(bottom) for configurations $AF_z$ и $F_z$ (curves 1 and 2 correspondingly).
Curve 3 is the difference of energies of these configurations. All
values $T_c$, $t_z,~t'$, $\D H$ are in unit $t$.

Fig.~2.
Density of state for two types $AF_z$ and $F_z$ of spin alignment
for models with small $t_z, t'$. Thin solid and dashed lines refer to
contributions in DOS from bonding and antibonding bands. The calculated DOS
is smoothed with gaussian function with dispersion $\d E=0.02t$, E is in
unit $t$.

Fig.~3.
Changes in the magnetic excitations branches $\o(q)$ during transformation
$F_z\to AF_z$ of spin configuration of model (5) with anisotropic
interlayer interaction with parameters $g_1=0.05$, $g_{2}=0.15$.
A right figure is schematic presentation of the spin excitations in case of
the spiral or modulated incommesurate spin structure with
$Q_{\eta} =(\pi (1-\eta ), \pi )$, $\eta=0.1$.
Squares mark the supposed resonance frequencies in various
interpretations.

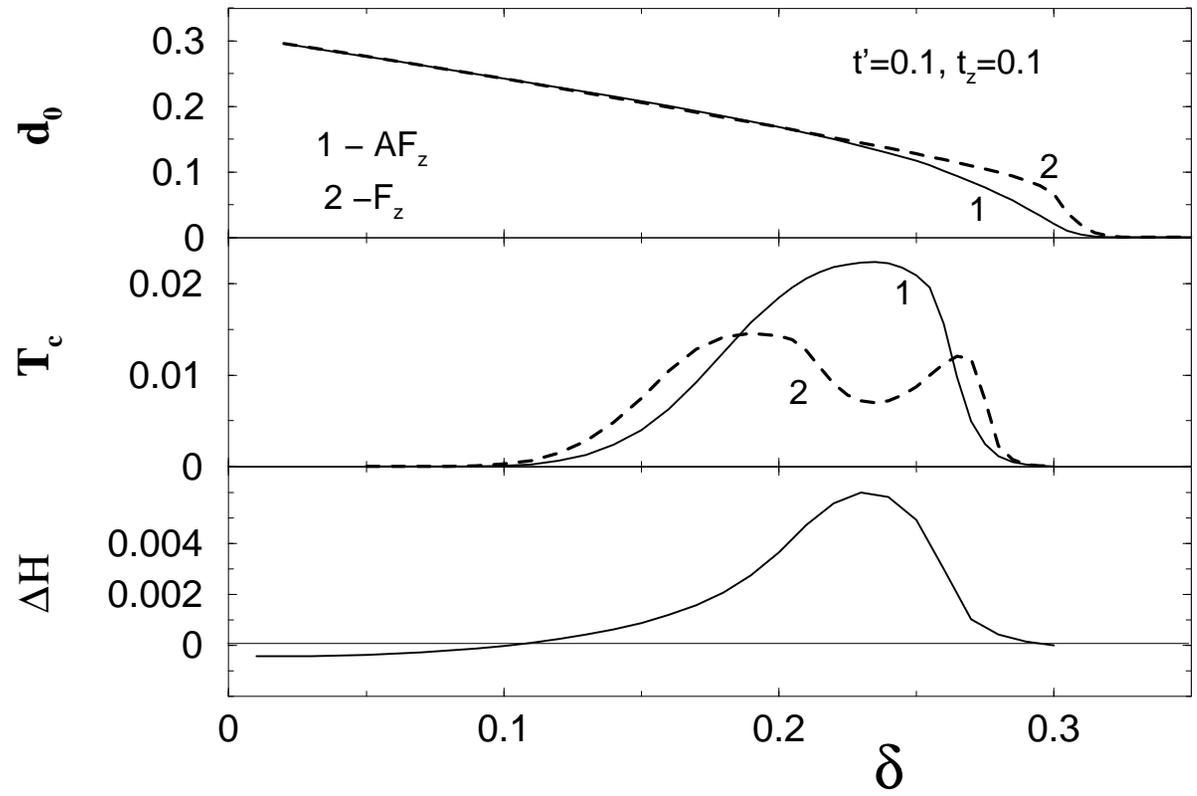
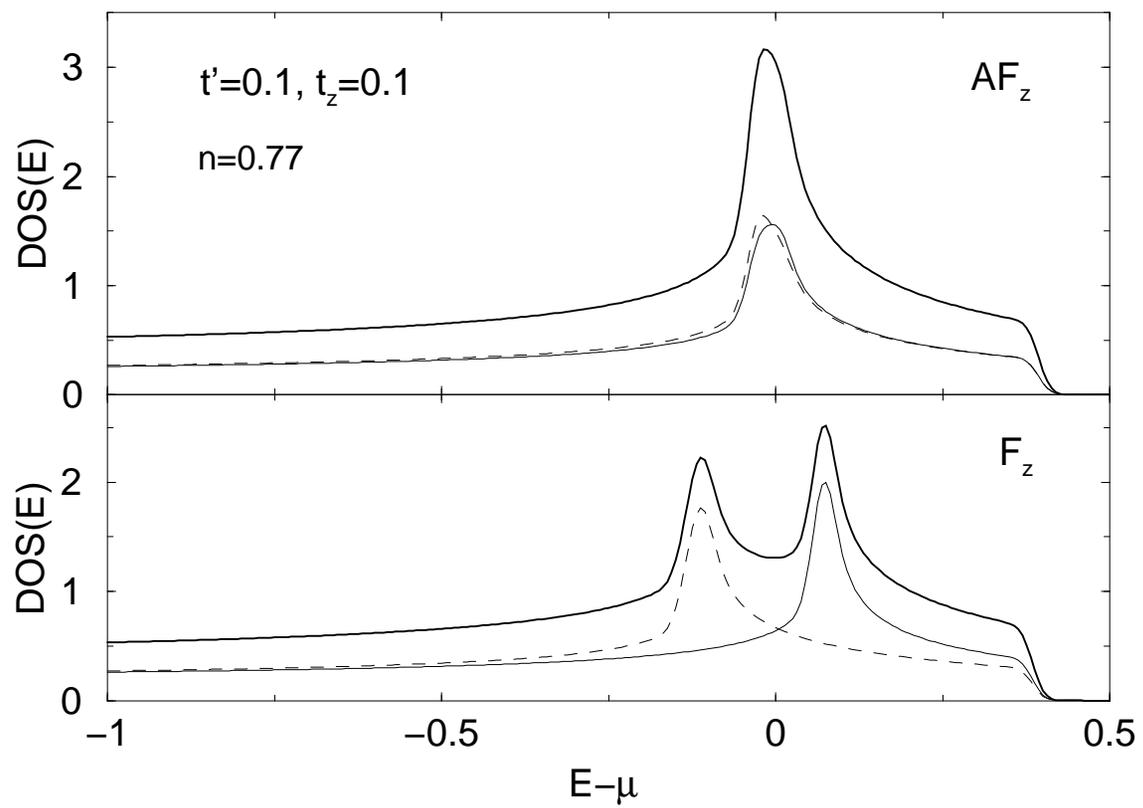
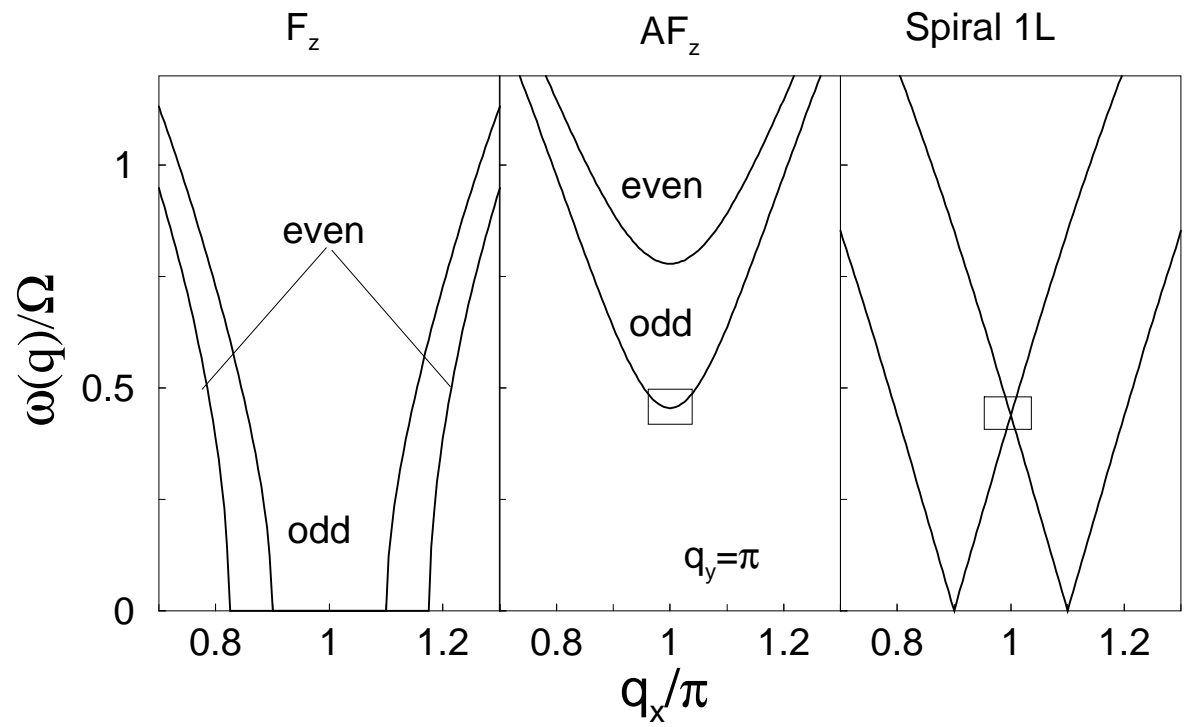
\end{document}